\documentclass[lettersize,journal]{IEEEtran}
\usepackage{amsmath,amsfonts}
\usepackage{algorithmic}
\usepackage{algorithm}
\usepackage{makecell}
\usepackage{array}
\usepackage{textcomp}
\usepackage{stfloats}
\usepackage{url}
\usepackage{hyperref}
\usepackage{verbatim}
\usepackage{graphicx}
\usepackage{cite}
\usepackage{xcolor}
\usepackage{multicol}
\usepackage{multirow}
\usepackage{tikz}
\usepackage{pgfplots}
\usepackage{pgfplotstable}
\pgfplotsset{compat=1.7}
\usetikzlibrary{patterns}
\usepackage{subcaption}

\begin{document}

\title{Code Refactoring with LLM: A Comprehensive Evaluation With Few-Shot Settings}

\author{Md. Raihan Tapader, Md. Mostafizer Rahman, Ariful Islam Shiplu, Md Faizul Ibne Amin, and Yutaka Watanobe
        % <-this % stops a space
%\thanks{This study was supported by the Japan Society for the Promotion of Science (JSPS) KAKENHI (Grant Number 19K12252).}% <-this % stops a space
\thanks{Md. Raihan Tapade is with the Department of Computer Science and Engineering, Dhaka University of Engineering \& Technology, Gazipur, Bangladesh (e-mail: raihantapader@gmail.com).}

\thanks{Md. Mostafizer Rahman is with the Department of Computer Science, Tulane University, New Orleans, LA, USA (e-mail: mostafiz26@gmail.com, mrahman9@tulane.edu).}

\thanks{Ariful Islam Shiplu is with the Department of Computer Science and Engineering, Dhaka University of Engineering \& Technology, Gazipur, Bangladesh (e-mail: shipluarifulislam@gmail.com).}

\thanks{ Md Faizul Ibne Amin is with the Department of Computer and Information Systems, University of Aizu, Japan (e-mail: aminfaizul007@gmail.com).}

\thanks{Yutaka Watanobe is with the Department of Computer and Information Systems, University of Aizu, Japan (e-mail: yutaka@u-aizu.ac.jp).}

%\thanks{\copyright 2025 IEEE.  Personal use of this material is permitted.  Permission from IEEE must be obtained for all other uses, in any current or future media, including reprinting/republishing this material for advertising or promotional purposes, creating new collective works, for resale or redistribution to servers or lists, or reuse of any copyrighted component of this work in other works.}

%\thanks{Manuscript received April 19, 2021; revised August 16, 2021.}
}


%\IEEEpubid{\copyright 2025 IEEE.  Personal use of this material is permitted.  Permission from IEEE must be obtained for all other uses, in any current or future media, including reprinting/republishing this material for advertising or promotional purposes, creating new collective works, for resale or redistribution to servers or lists, or reuse of any copyrighted component of this work in other works.}

\maketitle

\begin{abstract}
In today's world, the focus of programmers has shifted from writing complex, error-prone code to prioritizing simple, clear, efficient, and sustainable code that makes programs easier to understand. Code refactoring plays a critical role in this transition by improving structural organization and optimizing performance. However, existing refactoring methods are limited in their ability to generalize across multiple programming languages and coding styles, as they often rely on manually crafted transformation rules. The objectives of this study are to (i) develop an Large Language Models (LLMs) -based framework capable of performing accurate and efficient code refactoring across multiple languages (C, C++, C\#, Python, Java), (ii) investigate the impact of prompt engineering (Temperature, Different shot algorithm)  and instruction fine-tuning on refactoring effectiveness, and (iii) evaluate the quality improvements (Compilability, Correctness, Distance, Similarity, Number of Lines, Token, Character, Cyclomatic Complexity) in refactored code through empirical metrics and human assessment. To accomplish these goals, we propose a fine-tuned prompt-engineering-based model combined with few-shot learning for multilingual code refactoring. Experimental results indicate that Java achieves the highest overall correctness up to 99.99\% the 10-shot setting, records the highest average compilability of 94.78\% compared to the original source code and maintains high similarity ($\approx$53–54\%) and thus demonstrates a strong balance between structural modifications and semantic preservation. Python exhibits the lowest structural distance across all shots ($\approx$277–294) while achieving moderate similarity ($\approx$44–48\%) that indicates consistent and minimally disruptive refactoring. C\# consistently preserves semantic content, achieving up to 57.63\% similarity at 10-shot, despite larger structural changes. C and C++ show moderate distances ($\approx$266–338) and variable similarity (36–55\%), reflecting more unpredictable outcomes.
\end{abstract}


\section{Introduction}
\begin{figure}
    \centering
    \includegraphics[width=1\linewidth]{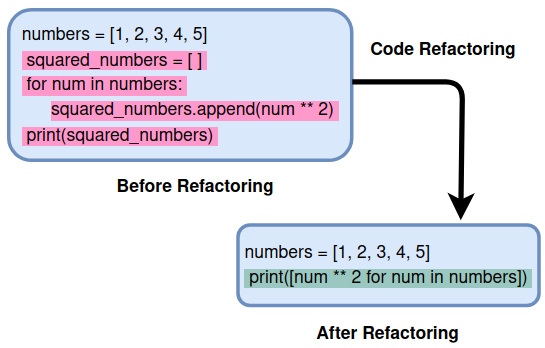}
    \caption{ A Readable Code Refactoring Example }
    \label{Code_Refactoring}
\end{figure}

Now-a-days, there is intense competition among programmers to write highly optimized code. The goal is to make code simple, concise, and easy to understand so that everyone from beginners to experts can read and maintain it effortlessly. Another major focus is on improving time and space efficiency. However, achieving this balance has become significant challenge because reducing execution time often increases memory uses, and vice versa. To address this, programmers continuously work on refactoring their code, improving earlier versions by removing redundancies, reducing the number of lines and tokens, and making the logic clearer. Code refactoring not only enhances performance but also improves code readability, maintainability, and adaptability to future changes as shown in Figure \ref{Code_Refactoring}. 

Over the few decades, code refactoring has evolved from being a purely manual and experience-driven practice to a systematic process supported by automated tools and frameworks. Early refactoring methods focused mainly on syntactic transformations, such as renaming variables, extracting methods, or eliminating duplicate code. Tools like Eclipse, IntelliJ IDEA, and other integrated development environments (IDEs) have helped standardize these transformation for making them accessible to a broader range of developers. However, as software systems have grown in size and complexity, the limitations of purely rule-based approaches have become apparent. Researchers have explored program analysis techniques such as abstract syntax trees (ASTs), control flow graphs (CFGs), and static analysis to detect code smells and suggest improvements. 

\begin{figure*}
    \centering
    \includegraphics[width=.9\linewidth]{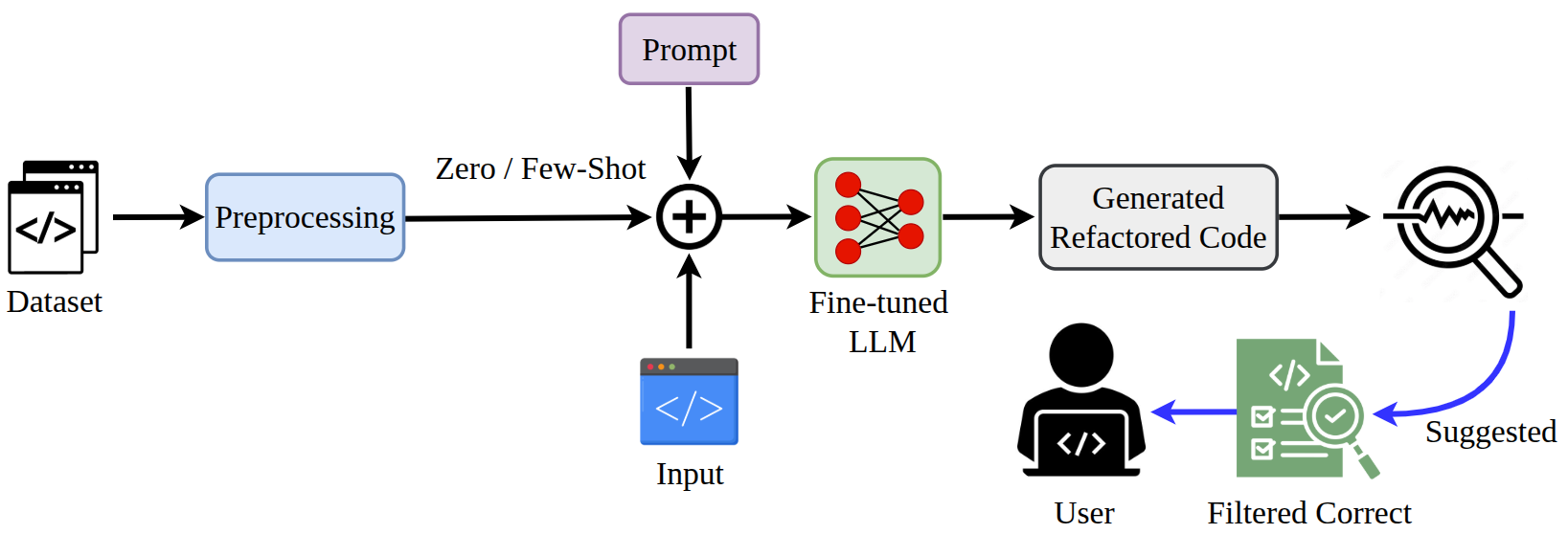}
    \caption{Proposed approach}
    
    \label{porposed method}
\end{figure*}

More recently, machine learning and deep learning models have been applied to learn refactoring patterns from large-scale repositories to move beyond the surface-level changes toward deeper structural and semantic optimization \cite{aniche2020effectiveness, sidhu2022machine, naik2024deep}. Despite showing promising potential, current automated refactoring models struggle to generalize across languages and coding styles because they are typically trained on limited, domain-specific datasets. Moreover many existing approaches primarily target easily detectable code smells \cite{kaur2016analysis} or syntactic issues, rather than capturing the deeper semantic intent and architectural considerations embedded in complex code. Additionally, the inherent complexity of source code, which involves hierarchical structures, control flow dependencies, and data flow relationships poses significant challenges for standard sequence-based or graph-based models to fully understand and process. 

Now, in the era of LLM, there is renewed interest in transforming how code refactoring is approached. Unlike earlier models that focused mainly on shallow syntactic changes or rule-based transformation, LLMs offer a deeper contextual understanding of both natural language and programming language \cite{rahman2025roberta, shirafuji2023program}. Studies \cite{shin2025prompt} have shown that prompt design significantly affects model performance in code-related tasks, while Chen et al. \cite{chen2021evaluating} demonstrated that natural language prompts can successfully instruct models like Codex to produce syntactically and semantically correct code. Pornprasit et al. \cite{pornprasit2024fine} use GPT-3.5 with zero-shot for code review automation and achieve impressive result. White et al. \cite{white2024chatgpt} work with Generative AI prompt pattern for code refactoring that help software engineering activities.

To the best of our knowledge, existing research does not adequately address several critical aspects of code refactoring. These include effectively highlighting key changes in refactored code, such as modifications in Variables and functions, estimating the success rates of refactoring operations, enabling seamless code transformation between different programming styles or paradigms, and ensuring bug-free correction of faulty code. Furthermore, there remains significant gap in supporting refactoring across diverse programming languages and in systematically exploring refactoring tasks that have yet to be thoroughly investigated. This lack of comprehensive tooling and methodological support ultimately limits programmer's and professional's ability to assess and improve code correctness, readability, maintainability, and security. 

\textbf{Scope}: This research focuses on developing a domain specific fine-tuned prompt engineering model, further integrated with specialized few-shot learning algorithms, and evaluate its effectiveness on a dataset spanning various programming languages. 

The key contributions of this paper are as follows:
\begin{itemize}

\item We have created a dataset that includes code written in five different programming languages (C, C++, C\#, Python, Java). In addition, we designed ten specialized prompts to guide the LLM in generating reliable code. 

\item Developed a model that combines fine-tuned prompt engineering with few-shot learning techniques to improve code refactoring performance.

\item Experimental result Shows that Java programming language achieved the best result 99.99\% correctness in 10-shot, and also achieved average 94.78\% compilability compared with source code.

\end{itemize}

\section{Related Works}

Recent advances in LLMs have significantly transformed automated code generation tasks. Several studies have explored prompt engineering as an effective approach to guide LLMs in producing syntactically and semantically correct code. Work such as \cite{shin2025prompt, cordeiro2024empirical} demonstrated that carefully crafted prompts can substantially improve model outputs without extensive retraining. However, these approaches often rely on manually designed prompts and do not address domain-specific adaptation. 

In parallel, few-shot learning techniques have emerged to reduce the need for large labeled datasets by enabling models to generalize from a limited number of examples \cite{shirafuji2023refactoring}. Prior research has shown promising results in applying few-shot learning to programming tasks. Yet most studies focus on single-language datasets or specific programming paradigms which is limit their applicability to diverse real-world environments. 

Moreover, while instruction fine-tuning has gained attention for aligning LLMs with specific tasks, its potential in the domain of code refactoring remains underexplored. Studies such as \cite{li2023instructcoder, zhang2025unveiling, yuan2023evaluating, ma2024llamoco} show that instruction-tuned models achieve better performance in code-related tasks (code editing, code generation) when explicitly guided by structured instructions. However, there is limited empirical research combining instruction fine-tuning with prompt engineering to assess their joint impact on code quality improvement across multiple languages.

In the context of code refactoring, recent works have proposed neural approaches that attempt to automate code transformation. While these models offer performance improvements over rule-based systems, they often lack explainability, generalizability, and  especially in multilingual settings. Metrics like cyclomatic complexity, code similarity, and code readability have been utilized to evaluate refactoring outcomes, but human evaluation remains essential to assess semantic preservation and maintainability.

Our work builds on this foundation by proposing an LLM-based multilingual refactoring framework, enriched with prompt engineering and instruction fine-tuning, and evaluates its effectiveness using both quantitative metrics (e.g., token length, character count, code similarity, cyclomatic complexity) and human-centered assessments.

\section{Methodology}

In this section, we provide a detailed description of our automated code refactoring model, which combines fine-tuned prompt engineering with few-shot learning to improve accuracy, efficiency, and generalizability. Leveraging prompt engineering allows the model to generate precise, context-aware code transformations, while few-shot learning facilitates adaptation to multiple programming languages and coding styles with minimal training data. The overall framework is depicted in Figure \ref{porposed method}.

\subsection{Overview}
\begin{figure}[!t]
    \centering
    \includegraphics[width=.9\linewidth]{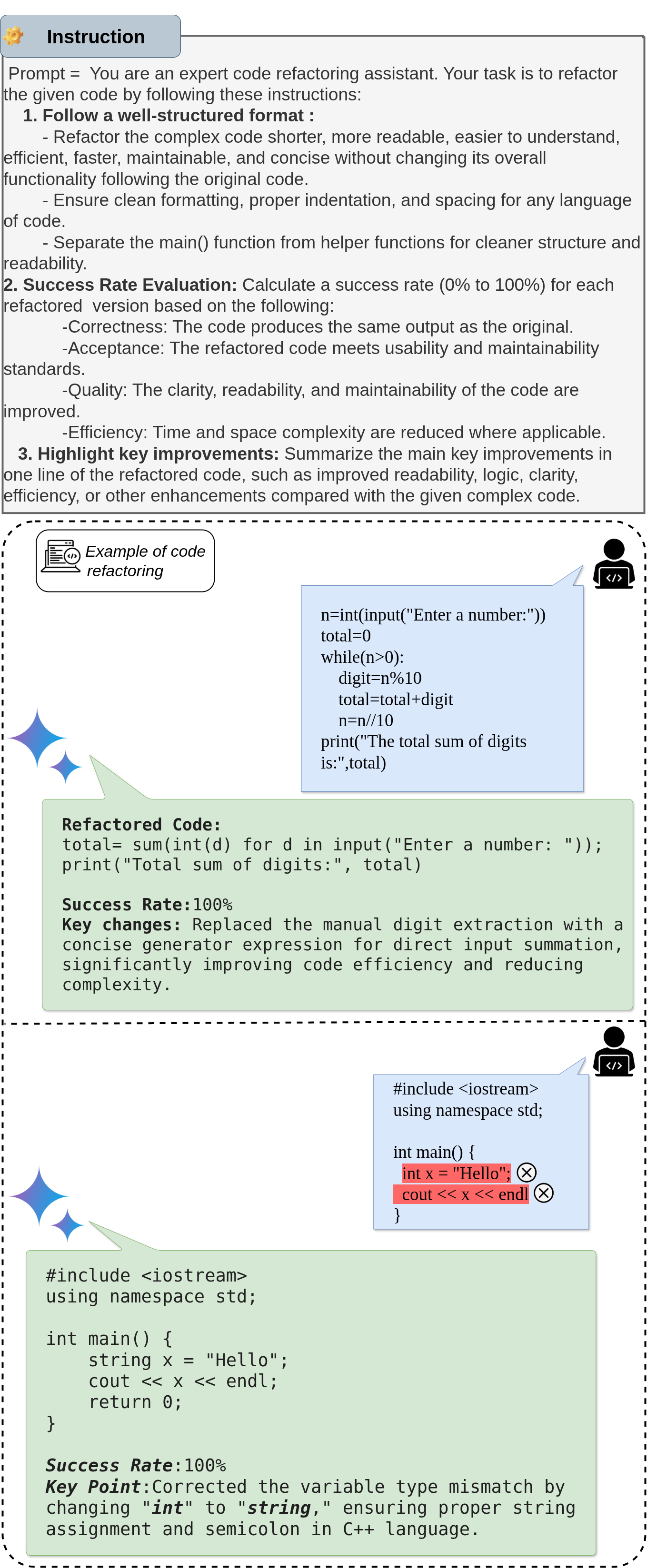}
    \caption{Demonstration of model behavior: (1) instruction interpertation, (2) refactoring responses to input program based on system instructions, and (3) error handling proficiency when refactoring erroneous program. The model exhibits excellent performance.}
    \label{Refactoring example}
\end{figure}

In recent years, extensive research by Shirafuji et al. \cite{shirafuji2023refactoring} has focused on improving refactoring techniques, exploring various methods to boost readability, optimize performance, and reduce complexity. Advanced large language models (LLMs), like OpenAI Codex, are being utilized to experiment with sophisticated refactoring approaches. These models have demonstrated the ability to improve code organization, clarity, and efficiency, enabling developers to produce higher-quality, more efficient code. Figure \ref{Refactoring example} presents a real instance of refactored code produced by the model is demonstrated to showcase its practical effectiveness.

\begin{figure}[h]
    \centering
    \includegraphics[width=1\linewidth]{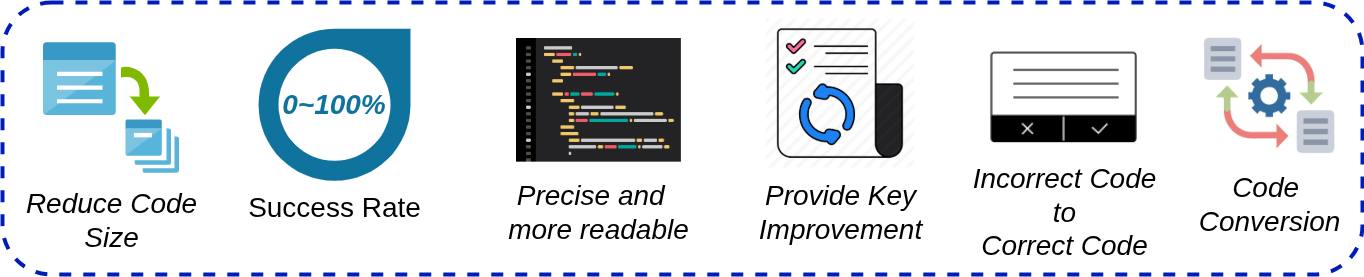}
    \caption{Showing the special features of code refactoring which used to design prpmpt}
    \label{Special_feature}
\end{figure}

In this study, we first define the prompting module by designing ten distinct prompts. Each prompt consists of three key components: a custom instruction that highlights specific features as illustrated in Figure \ref{Special_feature}, along with a complex code snippet and its corresponding refactored version. Next, we outline the prompting strategies employed in our methodology, which include zero-shot, two-shot, four-shot, and few-shot prompting. In zero-shot prompting, only the instruction is provided without any accompanying examples. Conversely, two-shot, four-shot, and few-shot prompting incorporate two, four, and several relevant examples respectively, alongside the instruction. During the learning phase, the input code is supplied to a large language model (e.g., GPT-3.5-turbo) together with few-shot examples for instruction fine-tuning. The model learns the transformation patterns from the prompts and examples and produces a newly fine-tuned model specifically optimized for multilingual code refactoring. Finally, the fine-tuned model generates multiple refactored versions of the input code. These outputs undergo rigorous validation through online compilers (e.g., Programiz), integrated development environments (IDEs) such as Visual Studio and Code::Blocks, as well as human evaluation to ensure syntactic correctness, semantic preservation, and enhanced maintainability. This workflow is further outlined in Algorithm \ref{algo}. 

\subsection{Dataset}
\begin{figure}[http]
    \centering
    \includegraphics[width=1\linewidth]{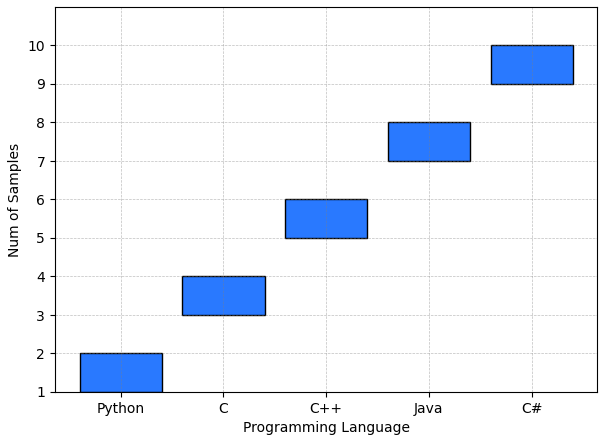}
    \caption{Dataset Samples Distribution}
    \label{Data samples}
\end{figure}

We curated a dataset \(\mathcal{D}\) consisting of 10 code samples from five programming languages: C, C++, C\#, Python, and Java. Each sample contains an \texttt{Instruction}, a \texttt{Complex Code} snippet, and its corresponding \texttt{Refactored Code}. The dataset includes 2 samples each for C and C++, 3 for Python, and 1 each for C\# and Java. Figure \ref{Data samples} depicts distribution of dataset samples. The instructions were systematically designed to ensure clarity, consistency, and relevance for refactoring tasks.

\begin{figure} [ht]
    \centering
    \includegraphics[width=1\linewidth]{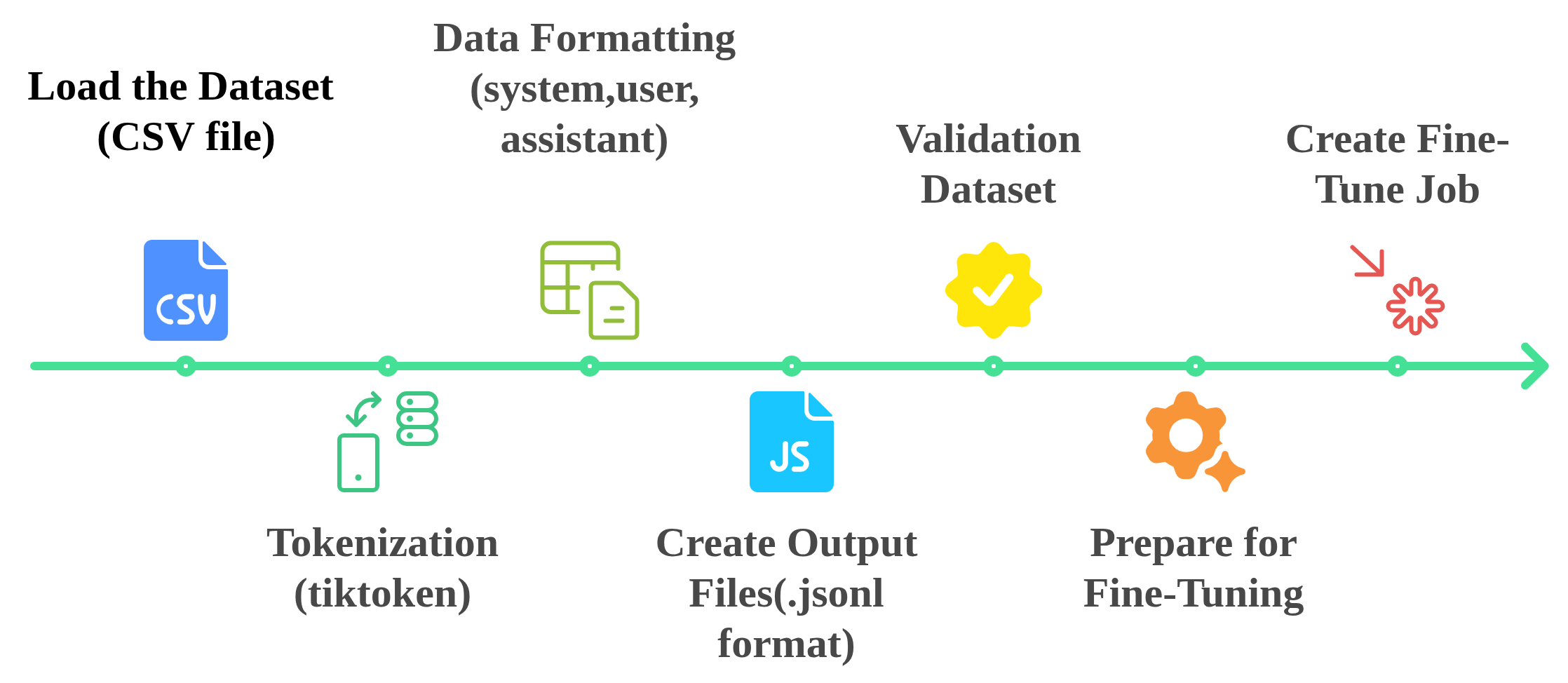}
    \caption{End-to-End pipeline for data preprocessing and model fine-tuning}
    \label{preprocessing}
\end{figure}

Raw data in CSV format was converted into OpenAI’s JSONL format with roles \texttt{system}, \texttt{user}, and \texttt{assistant}. We rigorously validated the dataset to exclude malformed entries, those exceeding token limits, or missing fields, using the \texttt{tiktoken} library to enforce a 4096-token maximum. The preprocessing pipeline is depicted in Figure \ref{preprocessing}.

\begin{algorithm}  \caption{LLM-Based Multilingual Code Refactoring with Prompt Design and Few-Shot Learning} \label{algo}
\begin{algorithmic}[1] \STATE \textbf{Input:} Annotated code dataset with instructions, complex code, and refactored code \STATE \textbf{Output:} Multiple refactored code versions with success metrics from fine-tuned LLM

\STATE \textbf{Dataset Preparation:} Load and convert dataset to OpenAI JSONL format; discard invalid or oversized examples

\STATE \textbf{Prompt Engineering:} Design zero-shot to few-shot prompts with selected example pairs embedded alongside the test code

\STATE \textbf{Model Fine-Tuning:} Fine-tune GPT-3.5-Turbo on prepared dataset to obtain model $M$

\STATE \textbf{Inference:} For each input code, generate prompts and query $M$ to produce multiple diverse refactored outputs

\STATE \textbf{Evaluation:} Assess each refactoring for Correctness, Cyclomatic Complexity, Line of Code, Compilability, Chars, Tokens, Similarity, Distance, and efficiency; summarize key improvements

\STATE \textbf{Validation:} Verify refactored code via online compilation, IDE testing, and human review \end{algorithmic} \end{algorithm}

\subsection{Refactoring Evaluation and Prompting Strategy}

The dataset \(\mathcal{D} = \{\text{Instruction}, \text{Complex Code}, \text{Refactored Code}\}\) supports evaluation via few-shot prompting. The number of examples \(n\) in the prompt varies by strategy:

\[
n = 
\begin{cases}
0 & \text{Zero-shot} \\
2 & \text{Two-shot} \\
4 & \text{Four-shot} \\
\vdots & \text{Few-shot}
\end{cases}, \quad n \in \{0, \dots, 10\}
\]

For each complex code snippet \(C_i\) in the set \(\mathcal{P} = \{C_1, C_2, \dots, C_N\}\), the prompt \(P_i\) is constructed as:

\[
P_i = \{\mathcal{I}_i, \mathcal{E}_i, C_i\}
\]

where \(\mathcal{I}_i\) is the instruction and \(\mathcal{E}_i = \{(C^{(k)}, R^{(k)})\}_{k=1}^n\) are example pairs of complex and refactored code. The fine-tuned GPT-3.5-turbo model \(M\) generates \(k=5\) refactoring candidates per prompt input:

\[
\mathcal{R}_i = M(P_i, \text{temperature}=t, n=k) = \{R_{i1}, \dots, R_{ik}\}
\]

\subsection{Model Configuration}
We utilize the GPT-3.5-turbo-0125 model, leveraging its strong foundation in instruction tuning and code understanding derived from Codex and RLHF.

The temperature parameter is tuned between 0.1 and 1.5 to balance creativity and consistency, with 0.3 yielding optimal performance. The maximum token limit for output generation is set to 2,048, accommodating average input lengths of approximately 110 tokens.

A unified system instruction guides the model to refactor code to be shorter, more readable, efficient, and maintainable without altering functionality. Key directives include:
\begin{itemize}
    \item Producing multiple distinct refactoring variants per input.
    \item Maintaining clean formatting, proper indentation, and descriptive naming.
    \item Separating the main function from helper functions for modularity.
    \item Evaluating refactorings on correctness, usability, quality, and efficiency.
    \item Summarizing key improvements succinctly for each candidate.
\end{itemize}

All inputs instruction, complex code, and refactored code are formatted as code blocks enclosed in triple backticks to maintain prompt clarity.

\section{Experimental Results} 
In this section, we present the evaluation metrics employed to assess the quality of the refactored code and discuss the experimental results obtained from applying our proposed approach.

\subsection{Matrics}

\subsubsection{\text{Pass@}k}

is a metric for evaluating the \textit{functional correctness} of generated programs by estimating the probability, introduced by Chen et al.\cite{chen2021evaluating}. It is defined in Equation \ref{eq:passatk}, where \(n\) is the total samples, \(c\) is the correct samples, and \(k\) is the evaluated samples, satisfying \(n \geq k\) and \(c \leq n\) rules. Specifically, Pass@1 measures how often the first output is correct, while Pass@10 checks if at least one of the top 10 outputs is correct \cite{shirafuji2023refactoring}.

\begin{equation}
\text{Pass@}k := \mathbb{E}\left[1 - \frac{\binom{n - c}{k}}{\binom{n}{k}} \right]
\label{eq:passatk}
\end{equation}

\subsubsection{\text{Correct@}k}

In this paper, we create \textit{Correct@k} metric as the ratio of correct outputs among the top-\(k\) generated results using few-shot prompting, as shown in Equation \ref{eq: refactoring condition}. This metric is specifically designed to assess the correctness of the \(k\) generated samples, where \(n\) represents the total number of generated outputs, \(e\) denotes the number of erroneous ones, and \((n-e)\) indicates the number of correct outputs. The parameter \(k\) controls the level at which correctness is measured.

\begin{equation}
\text{Correct}@k =
\begin{cases}
100\%, & k \leq (n-e),\\
\frac{n-e}{k}, & k > (n-e).
\end{cases}
\label{eq: refactoring condition}
\end{equation}

For each new input code snippet \(C_i\), we generate \(n = 5\) refactored versions and evaluate  correctness using \textit{Correct@k}, where \(k \in \{1, 2, \dots, 5\}\). In this work, Correct@1 refers to the proportion in which at least one of the generated outputs is correct, whereas Correct@5 reflects the cases where all five outputs are correct. Subsequently, if a generated code \(R_i\) satisfies the correctness criteria defined by the equation, the success rate is computed accordingly.

\subsubsection{McCabe's Cyclomatic Complexity(CC)}
is a software metric used to measure the number of linearly independent paths through a program’s source code. It quantifies the program complexity by counting the number of decision points(e.g., if statements, while loops, for loops, switch statements)  in the code \cite{mccabe1976complexity}. A lower CC is preferable and makes the code easier to understand and maintain in software development. We employed a custom AST-based  library to accurately compute the CC for both correct and incorrect code samples.

\subsubsection{Compilability}
refers to whether a generated program is syntactically correct and successfully compiles, reflecting the model’s capabilities to produce valid and executable code It is calculated as equation \ref{eq:compilability}: 

\begin{equation}
\text{Compilability} = \Big( \frac{P}{P + E}\Big) \times 100\%
\label{eq:compilability}
\end{equation}

where P denotes the number of programs that compile successfully, and E denotes the number of programs that fail to compile. In our case, the comparability of the complex(original) programs used for refactoring is 100\%, as all of them were correct and problem-solved.

\subsubsection{Lines of Code (LOC)} is a software metric that measures the number of lines in a source code. In this work, we use Source Lines of Code (SLOC), which excludes comments, blank lines, and structural braces (e.g.,\{ \}), focusing solely on the actual executable statements. For refactoring analysis, SLOC is preferred over raw LOC because it avoids inflated counts from non-functional lines and gives a more accurate picture of the actual number of lines and logic implemented rather than comments or empty lines.

\subsubsection{Chars} refers to the total length of how much character is used in a code on average. This matrix sometimes yields anomalous values due to long string literals, comments, and excessive whitespace. To accurately measure the character length, we normalize the refactored code by removing such anomalies. For example, consider the string literals:

A='Enter a num:' and B= "H"; these would initially result in different character counts. However, after normalization, both are treated equivalently as A = ' ' and B = " ", leading to consistent and fair character length comparisons.

\subsubsection{Tokens} refers to the total length of how many tokens are used in a code on average. Tokens are the fundamental components of a program that the compiler processes. In this work, we used an official Python tokenizer for Python code, while for other languages, we employed a regex-based tokenizer.  To ensure accurate token counts, all code snippets are normalized by removing comments, newlines, and indentation before tokenization.

\begin{table*}[!t]
\caption{Quantitative comparison of Compilability, CC, LOC, character count, and token count for refactored programs across five programming languages under varying shot settings.}
\label{T_code}
\centering
\begin{tabular}{c|c|c|c|c|c|c}
\hline
\textbf{Shots} & \textbf{Programming Language} & \textbf{Compilability} & \textbf{CC} & \textbf{LOC} & \textbf{Chars} & \textbf{Tokens} \\ \hline \hline

\multirow{6}{*}{\centering 0-shot} 
& C      & 90.00 & 4.99 (±0.07) & 6.04 (±0.83)  & 197.56 (±11.40) & 150.25 (±9.41) \\ \cline{2-7}
& C++    & 92.67 & 4.94 (±0.27) & 8.15 (±1.51)  & 232.13 (±18.13) & 156.36 (±14.71) \\ \cline{2-7}
& C\#    & 88.67 & 5.01 (±0.27) & 6.13 (±2.19)  & 223.08 (±12.71) & 146.61 (±9.74) \\ \cline{2-7}
& Python & 91.33 & 2.10 (±0.97) & 6.06 (±0.48)  & 287.89 (±26.79) & 123.71 (±6.62) \\ \cline{2-7}
& Java   & 90.00 & 3.32 (±1.10) & 8.34 (±3.16)  & 229.19 (±39.94) & 118.69 (±31.99) \\ 

\hline
\multirow{5}{*}{\centering 2-shot} 
& C      & 83.00 & 4.95 (±0.46) & 6.17 (±1.32)  & 217.43 (±17.18) & 168.79 (±16.92) \\ \cline{2-7}
& C++    & 90.00 & 4.99 (±0.17) & 7.51 (±1.10)  & 238.74 (±39.69) & 165.69 (±18.37) \\ \cline{2-7}
& C\#    & 94.00 & 4.99 (±0.07) & 11.02 (±1.23) & 255.74 (±9.04)  & 164.53 (±5.59)  \\ \cline{2-7}
& Python & 94.67 & 2.11 (±0.54) & 7.23 (±0.98)  & 246.65 (±24.43) & 123.77 (±6.11)  \\ \cline{2-7}
& Java   & 94.00 & 4.92 (±0.62) & 8.59 (±1.10)  & 323.83 (±32.12) & 185.90 (±24.52) \\ \hline

\multirow{5}{*}{\centering 4-shot} 
& C      & 87.00 & 4.88 (±0.33) & 7.43 (±2.21)  & 229.91 (±19.59) & 156.73 (±15.23) \\ \cline{2-7}
& C++    & 92.67 & 4.97 (±0.13) & 9.17 (±1.98)  & 265.78 (±16.64) & 160.19 (±9.54)  \\ \cline{2-7}
& C\#    & 97.33 & 4.94 (±0.18) & 11.64 (±1.28) & 269.29 (±13.04) & 161.45 (±7.86)  \\ \cline{2-7}
& Python & 93.33 & 1.79 (±1.05) & 6.47 (±1.20)  & 269.77 (±28.52) & 120.05 (±7.73)  \\ \cline{2-7}
& Java   & 93.33 & 4.37 (±0.89) & 8.34 (±1.60)  & 300.25 (±46.57) & 167.35 (±34.74) \\ 
\hline

\multirow{5}{*}{\centering 6-shot} 
& C      & 93.00 & 4.91 (±0.38) & 12.29 (±2.58) & 252.03 (±20.88) & 159.07 (±16.49) \\ \cline{2-7}
& C++    & 94.00 & 2.58 (±0.68) & 9.89 (±1.62)  & 223.39 (±24.46) & 112.15 (±15.79) \\ \cline{2-7}
& C\#    & 96.67 & 4.94 (±0.22) & 10.84 (±1.60) & 273.36 (±16.97) & 163.35 (±12.40) \\ \cline{2-7}
& Python & 87.33 & 2.50 (±1.16) & 7.29 (±1.71)  & 237.57 (±43.22) & 121.39 (±10.13) \\ \cline{2-7}
& Java   & 96.00 & 2.33 (±0.62) & 6.67 (±1.55)  & 190.10 (±28.23) & 83.93 (±20.01)  \\ 
\hline

\multirow{5}{*}{\centering 8-shot} 
& C      & 93.33 & 4.95 (±0.25) & 11.67 (±3.19) & 260.99 (±26.76) & 160.63 (±14.55) \\ \cline{2-7}
& C++    & 91.33 & 2.93 (±1.10) & 9.92 (±2.01)  & 225.71 (±42.30) & 116.18 (±27.93) \\ \cline{2-7}
& C\#    & 95.33 & 4.73 (±0.49) & 9.20 (±1.53)  & 272.13 (±28.72) & 162.07 (±21.53) \\ \cline{2-7}
& Python & 87.33 & 1.88 (±1.08) & 5.84 (±1.65)  & 294.47 (±36.28) & 120.63 (±10.53) \\ \cline{2-7}
& Java   & 96.67 & 4.99 (±0.07) & 8.97 (±1.59)  & 344.15 (±14.53) & 195.21 (±13.95) \\ 
\hline

\multirow{5}{*}{\centering 10-shot} 
& C      & 96.00 & 4.97 (±0.21) & 12.05 (±2.79) & 259.00 (±19.05) & 158.39 (±12.58) \\ \cline{2-7}
& C++    & 92.00 & 4.93 (±0.29) & 12.14 (±2.23) & 303.55 (±21.78) & 162.53 (±13.89) \\ \cline{2-7}
& C\#    & 95.33 & 4.91 (±0.29) & 10.91 (±2.42) & 307.29 (±32.90) & 162.00 (±15.84) \\ \cline{2-7}
& Python & 90.67 & 3.32 (±1.16) & 8.95 (±1.91)  & 271.87 (±31.99) & 112.81 (±9.98)  \\ \cline{2-7}
& Java   & 98.67 & 4.92 (±0.18) & 8.81 (±1.46)  & 344.39 (±18.07) & 190.32 (±15.01) \\ 
\hline

\end{tabular}
\end{table*}

\subsubsection{Levenshtein Distance} also known as edit distance, is a string metric that quantifies the difference between two sequences (typically strings) by counting the minimum number of single-character edits (insertions, deletions, or substitutions) required to transform one string into the other. In this study, we do not normalize the code before comparison in order to showing the actual distance bteween original and refactored code. For example, consider the following two strings:

A: "int x+y+z",  and B: "int y-z". The total number of edits needed to convert A into B is 3, so the Levenshtein distance is 3.

\subsubsection{Similarity} is a metric used to measure how alike the original, and refactored code are, based on the Levenshtein distance.  A higher distance implies lower similarity, and vice versa. In this study, Similarity between two programs a,b, with lengths \( \lvert a \rvert \), \( \lvert b \rvert \), is defined as Equation \ref{eq:similarity}: 

\begin{equation}
\text{Similarly}(a, b) = \left( 1 - \frac{\text{Distance}(a, b)}{\max(|a|, |b|)} \right)
\label{eq:similarity}
\end{equation}

Here, Distance (a,b) represents the Levenshtein distance between the two programs.

\subsection{Results}
We conducted comprehensive experiments across multiple programming languages using a fine-tuned large language model with few-shot prompting to assess its effectiveness in automated code refactoring. Table \ref{T_code} summarizes key code quality metrics—Compilability, CC, LOC, Chars, and token count for five programming languages (C, C++, C\#, Python, and Java) evaluated under different few-shot prompting scenarios ranging from 0-shot to 10-shot.

\begin{figure*}
\captionsetup{justification=centering}
     \centering
     \begin{subfigure}[b]{0.32\textwidth}
 \centering
  \includegraphics[width=1\linewidth]{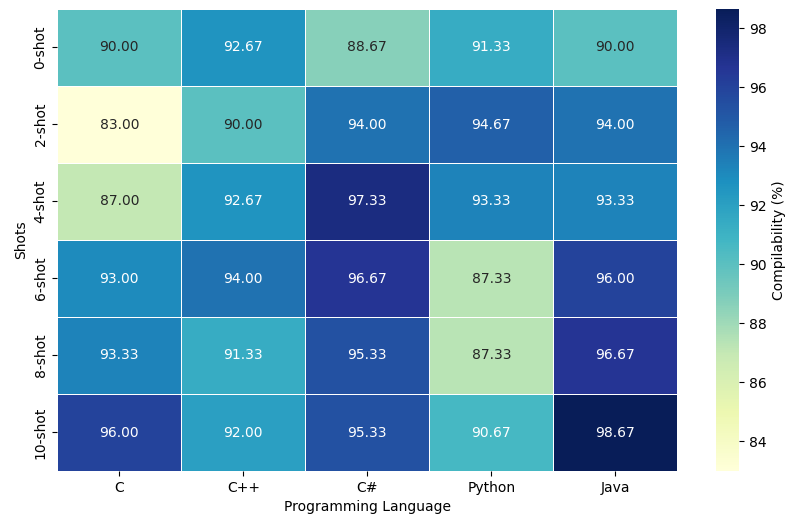}
        \caption{Compilability}
        \label{compilability}
     \end{subfigure}
     \hfill
       \begin{subfigure}[b]{0.32\textwidth}
     \centering
  \includegraphics[width=1\linewidth]{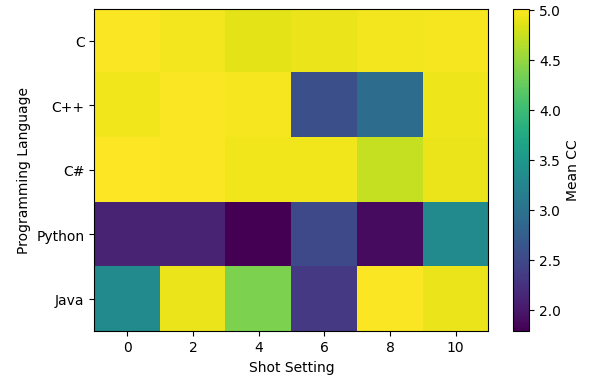}
        \caption{Cyclomatic Complexity}
        \label{cc}
     \end{subfigure}
     \hfill
    \begin{subfigure}[b]{0.32\textwidth}
     \centering
  \includegraphics[width=1\linewidth]{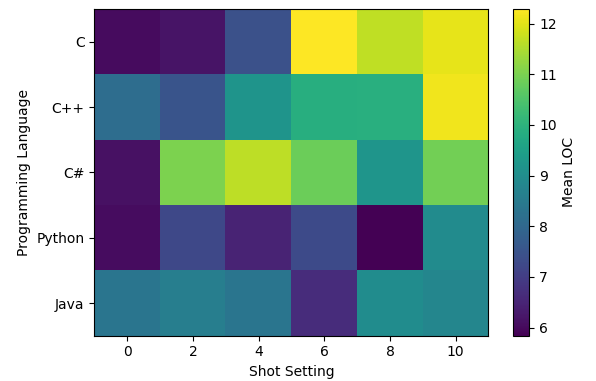}
        \caption{Lines of Code}
        \label{loc}
     \end{subfigure}
     \hfill

       \begin{subfigure}[b]{0.48\textwidth}  
       \centering
  \includegraphics[width=.7\linewidth]{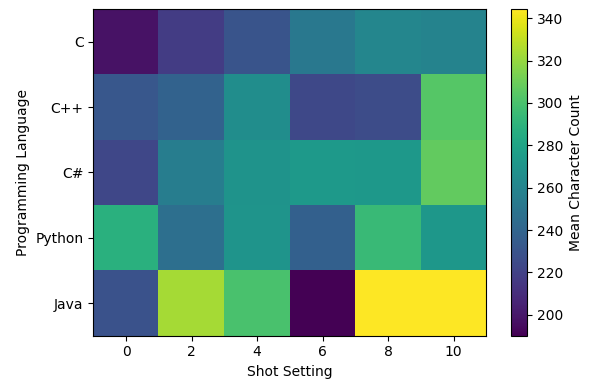}
        \caption{Number of Characters}
        \label{chars}
     \end{subfigure}
     \hfill
    \begin{subfigure}[b]{0.48\textwidth}
     \centering
  \includegraphics[width=.7\linewidth]{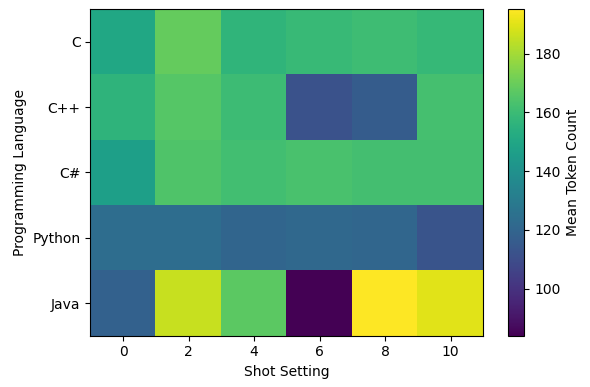}
        \caption{Number of Tokens}
        \label{tokens}
     \end{subfigure}
     \hfill

    \caption{Visualization of program metrics across five programming languages (C, C++, C\#, Python, Java) and varying shot settings. Subplots show (a) Compilability (\%), (b) Cyclomatic Complexity, (c) Lines of Code, (d) Number of Characters, and (e) Number of Tokens, highlighting trends and variations across 0-shot to 10-shot settings.}
    \label{all_metrics}
\end{figure*}

In the 0-shot baseline, compilability ranges narrowly between 88.67\% (C\#) and 92.67\% (C++) which indicates that even without examples, the model maintains reasonable syntactic consistency. CC is relatively stable for C, C++, and C\# around 5.0, while Python and Java generate simpler control flows with lower complexity 2.10 and 3.32, respectively. The refactored programs at this stage are compact, mostly between 6–8 lines of code and 200–290 characters, with Python being the most concise in token count 123.71.
Introducing two example prompts significantly alters the compilability trends across languages. While C experiences a decline to 83\% compilability, higher-level managed languages such as C\#, Python, and Java demonstrate notable improvements, reaching compilability rates of 94\% and above. Concurrently, LOC increase modestly for all languages, with particularly pronounced growth in C\# (11.02 lines) and Java (8.59 lines), reflecting longer and potentially more detailed function implementations. Corresponding increases are also observed in character and token counts, consistent with the enhanced code elaboration.
At the 4-shot prompting level, compilability stabilizes across languages, with C\# achieving the highest rate of 97.33\%. Python’s CC further decreases to 1.79, indicating the generation of highly linear and simplified code structures. Meanwhile, C++ and Java exhibit increased verbosity, each surpassing 260 characters on average. C\# continues to show the greatest increase in LOC, reaching 11.64 lines, suggesting that the model tends to produce more descriptive or boilerplate code when provided with additional contextual examples.
A notable shift in trends emerges at the 6-shot prompting level. Java attains a compilability of 96\%, closely followed by C\# at 96.67\%. In contrast, Python experiences a significant decline to 87.33\%, reversing its previous stability. Interestingly, C++ exhibits a marked reduction in cyclomatic complexity to 2.58, indicating that additional examples encourage the model to generate code with fewer branching paths. Meanwhile, C reaches its highest lines of code count at 12.29, suggesting that the model begins producing more extended and detailed low-level code structures at this stage.
At 8-shot, performance remains high for all except Python still 87.33\%. Java reaches 96.67\%, while C and C\# maintain 95\%. Complexity levels remain low for Python 1.88 and C++ 2.93. Java again produces the most verbose outputs, reaching 344.15 characters and 195.21 tokens, suggesting that the model begins over-eliciting large output structures.
The best compilability across the entire table is achieved here, with 98.67\% for Java and 96.00\% for C, while all other languages score above 92\%. CC values converge toward 5.0 for C, C++, and C\#, whereas Python and Java retain moderate complexity. LOC increases further, with C++ (12.14) and C\# (10.91) yielding longer functions, while Python remains compact 8.95 LOC. Token count reaches its overall peak in Java 190.32 tokens, reinforcing that Java scales in verbosity proportionally with more examples. Figure \ref{all_metrics} shows the heatmap visualization of all metrics.

\begin{comment}

\begin{figure*}
    \centering

    \begin{subfigure}[b]{0.32\textwidth}
        \centering
        \includegraphics[width=.9\linewidth]{compilabilityHeatMap.png}
        \caption{Compilability}
        \label{compilability}
    \end{subfigure}
    \hfill
    \begin{subfigure}[b]{0.32\textwidth}
        \centering
        \includegraphics[width=.9\linewidth]{cyclomaticComplexity Heatmap.png}
        \caption{Cyclomatic Complexity}
        \label{cc}
    \end{subfigure}

  \hfill

    \begin{subfigure}[b]{0.32\textwidth}
        \centering
        \includegraphics[width=.9\linewidth]{LOCHeatMap.png}
        \caption{Lines of Code}
        \label{loc}
    \end{subfigure}
    \hfill
    \begin{subfigure}[b]{0.45\textwidth}
        \centering
        \includegraphics[width=\textwidth]{CharsHeatMap.png}
        \caption{Number of Characters}
        \label{chars}
    \end{subfigure}

    \vspace{0.5cm}

    \begin{subfigure}[b]{0.45\textwidth}
        \centering
        \includegraphics[width=\textwidth]{TokenHeatMap.png}
        \caption{Number of Tokens}
        \label{tokens}
    \end{subfigure}

    \caption{Visualization of program metrics across five programming languages (C, C++, C\#, Python, Java) and varying shot settings. Subplots show (a) Compilability (\%), (b) Cyclomatic Complexity, (c) Lines of Code, (d) Number of Characters, and (e) Number of Tokens, highlighting trends and variations across 0-shot to 10-shot settings.}
    \label{all_metrics}
\end{figure*}

\end{comment}

\begin{table*}[!t]
\caption{Quantitative comparison between original and LLM-refactored programs across five programming languages, evaluated on Compilability, Cyclomatic Complexity, Lines of Code, character count, and token count. Each pair of rows reports the baseline properties of the original implementation and the corresponding refactored variant generated by the model.}
\label{compare}
\centering
\begin{tabular}{c|c|c|c|c|c|c}
\hline
\textbf{Programming Language} & \textbf{\#} & \textbf{Compilability} & \textbf{CC} & \textbf{LOC} & \textbf{Chars} & \textbf{Tokens} \\ \hline \hline

\multirow{2}{*}{\centering C} 
& Original      & 99.99 & 5.00 (± 0.00) & 19 (± 0.00)  & 387 (± 0.00) & 176 (± 0.00) \\ \cline{2-7}
& Refactored    & 90.39 & 4.94 (± 0.28) & 9.28 (± 2.15)  & 236.15 (± 19.14) & 158.98 (± 14.20) \\

\hline
\multirow{2}{*}{\centering C++} 
& Original      & 99.99 & 5.00 (± 0.00) & 20 (± 0.00)  & 419 (± 0.00) & 185 (± 0.00) \\ \cline{2-7}
& Refactored    & 92.11 & 4.22 (± 0.44) & 9.46 (± 1.74)  & 248.22 (± 27.17) & 145.52 (± 16.71) \\  \hline

\multirow{2}{*}{\centering C\#} 
& Original      & 99.99 & 5.00 (± 0.00) & 20 (± 0.00)  & 426 (± 0.00) & 181 (± 0.00) \\ \cline{2-7}
& Refactored    & 94.56 & 4.92 (± 0.25) & 9.96 (± 1.71)  & 266.81 (± 18.90) & 160.00 (± 12.16)  \\ \hline

\multirow{2}{*}{\centering Python} 
& Original      & 99.99 & 5.00 (± 0.00) & 18 (± 0.00) & 407 (± 0.00) & 135 (± 0.00) \\ \cline{2-7}
& Refactored    & 90.78 & 2.28 (± 0.99) & 6.97 (± 1.32)  & 268.04 (± 31.87) & 120.39 (± 8.52) \\ \hline

\multirow{2}{*}{\centering Java} 
& Original      & 99.99 & 5.00 (± 0.00) & 21 (± 0.00) & 487 (± 0.00) & 198 (± 0.00) \\ \cline{2-7}
& Refactored    & 94.78 & 4.14 (± 0.58) & 8.29 (± 1.74)  & 288.65 (± 29.91) & 156.90 (± 23.37) \\ \hline

\end{tabular}
\end{table*}

\begin{figure} [!t]
     \centering
         \begin{tikzpicture}[scale=1]
            \begin{axis}[
    ybar=.13cm,
    every node near coord/.append style={font=\tiny},
    legend style={font=\tiny},
    tick label style={font=\tiny},
    ylabel near ticks, ylabel shift={-6pt},
    %xlabel shift={-10pt},
    %width=\textwidth,
    width=6.8cm,
    height=4cm,
    every node near coord/.append style={
                        anchor=west,
                        rotate=75
                },
    enlargelimits=.15,
    enlarge y limits={0.1,upper},
    legend style={at={(0.5,-0.22)},
    anchor=north, legend columns=-1},
    ymin=30, 
    ylabel={$\mathbf{Compilability}$ (\%)},
    symbolic x coords={$C$, $C++$, $C\#$, $Python$, $Java$},
    xtick=data,
    %nodes near coords,
    ytick={10, 20, 30, 40, 50, 60, 70, 80, 90, 100},
    %x tick label style={rotate=25,anchor=east},
    grid=both,
    nodes near coords,
    nodes near coords align={vertical},
    bar width=8pt,
    %ymajorgrids=true,
    label style={font=\footnotesize},
    ]
%\addplot [draw=black, semithick, pattern=crosshatch, pattern color = black] coordinates {($C$,50.68) ($C++$,50.68) ($C\#$,49.32)};  % Macro F1-Score for 0.001

\addplot [draw=black, semithick, pattern=grid,  pattern color = black] coordinates {($C$,99.99) ($C++$,99.99) ($C\#$,99.99) ($Python$,99.99) ($Java$,99.99)};  % Macro F1-Score for 0.005

\addplot [draw=black, semithick, pattern=crosshatch dots,  pattern color = black] coordinates {($C$,90.39) ($C++$,92.11) ($C\#$,94.56) ($Python$,90.78) ($Java$,94.78)}; % Macro F1-Score for 0.01

%\addplot [draw=black, semithick, pattern=north east lines, pattern color = black] coordinates {($Sorting$,26326) ($Searching$,26173) };

\legend{$Original$, $Refactored$}
\end{axis}
 
        \end{tikzpicture}
        \caption{Compilability accuracy across original and refactored code}
         \label{compile}
     
\end{figure}
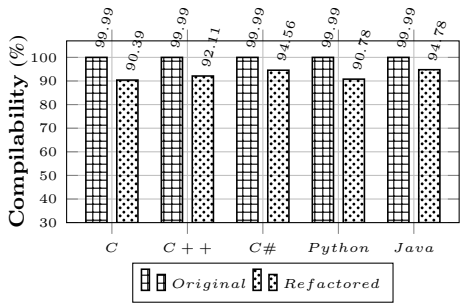

Table \ref{compare} presents a quantitative comparison between original and LLM-refactored code across five programming languages (C, C++, C\#, Python, and Java) evaluated on Compilability, CC, LOC, character count, and token count. Across all languages, the original code exhibits perfect or near-perfect compilability (99.99\%), serving as a strong ground truth baseline. In refactored versions, Java demonstrates the highest post-refactoring reliability (94.78\%), closely followed by C\# (94.56\%) which indicates that strongly-typed, object-oriented languages remain structurally resilient under automated transformations. Conversely, C and Python experience the greatest degradation, suggesting that syntax rigidity in C and formatting sensitivity in Python make them more vulnerable to syntactic inconsistencies during refactoring. In terms of Cyclomatic Complexity, refactored code tends to preserve or slightly reduce structural complexity across most languages. For example, C++ decreases from 5.00 to 4.22, and Java from 5.00 to 4.14, suggesting that the model often favors simpler control flow. Python shows the largest drop, from 5.00 to 2.28, implying that LLM refactoring may employ more linear or flattened logic structures in dynamic languages. A major change is observed in code length metrics. All refactored programs show a significant reduction in Lines of Code, with average LOC dropping from 18–21 lines to approximately 7–10 lines across all languages. This compression effect is further reflected in character count and token count, both of which decrease sharply. For instance, Java reduces from 487 to 288 characters, and Python from 407 to 268, while token counts drop from $\approx$180 down to $\approx$150 or lower. Interestingly, C\# and Java retain higher CC values while achieving moderate LOC reductions which suggestes that the model preserves structural richness while shortening syntax. Conversely, Python shows aggressive simplification both in structure and length, reflecting language-specific biases in refactoring behavior. Figure \ref{compile} illustrates the comparison between original and our proposed model performance.

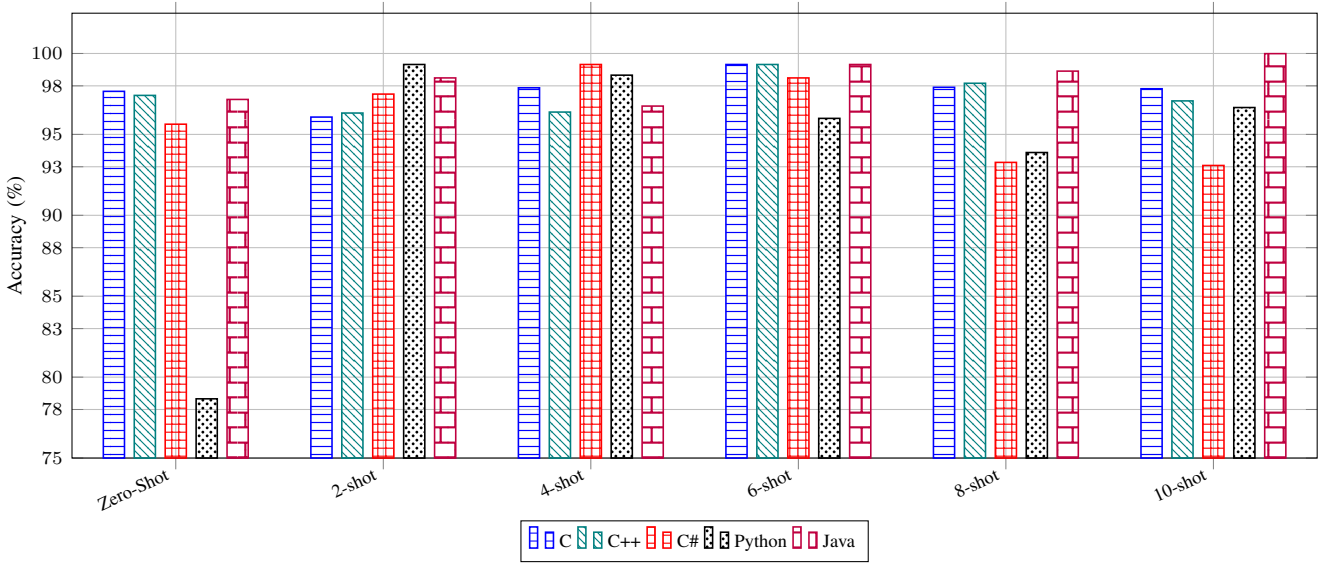
\begin{figure*}[!t]
     \centering
         \captionsetup{justification=centering}
         
         \begin{tikzpicture}[scale=1]
            \begin{axis}[
    ybar=.13cm,
    every node near coord/.append style={font=\scriptsize},
    legend style={font=\scriptsize},
    tick label style={font=\scriptsize},
    ylabel near ticks, ylabel shift={-5pt},
    %xlabel shift={-10pt},
    width=\linewidth,
    height=7.5cm,
    every node near coord/.append style={
                        anchor=west,
                        rotate=75
                },
    enlargelimits=.10,
    enlarge y limits={0.1,upper},
    legend style={at={(0.50, -0.14)},
    anchor=north,legend columns=-1},
    ymin=75,     
    ylabel={Accuracy (\%)},
    symbolic x coords={Zero-Shot, 2-shot, 4-shot, 6-shot, 8-shot, 10-shot},
    xtick=data,
    %nodes near coords,
    ytick={75,78,80,83,85,88,90,93,95,98, 100},
    x tick label style={rotate=25,anchor=east},
    grid=both,
    %nodes near coords,
    nodes near coords align={vertical},
    bar width=8pt,
    %ymajorgrids=true,
    label style={font=\footnotesize},
    ]
\addplot [draw=blue, semithick, pattern=horizontal lines, pattern color = blue] coordinates {(Zero-Shot, 97.66) (2-shot, 96.07) (4-shot, 97.88) (6-shot, 99.32 ) (8-shot, 97.91) (10-shot, 97.82)};  

\addplot  [draw=teal, semithick, pattern=north west lines,  pattern color = teal] coordinates {(Zero-Shot, 97.41) (2-shot, 96.32 ) (4-shot, 96.38) (6-shot, 99.32) (8-shot, 98.16) (10-shot, 97.07)}; 

\addplot [draw=red, semithick, pattern=grid, pattern color = red] coordinates {(Zero-Shot, 95.63) (2-shot, 97.49) (4-shot, 99.32) (6-shot, 98.49) (8-shot, 93.27) (10-shot, 93.08)}; 

\addplot [draw=black, semithick, pattern=crosshatch dots,  pattern color = black] coordinates {(Zero-Shot, 78.66) (2-shot, 99.32) (4-shot, 98.66 ) (6-shot, 95.99) (8-shot, 93.88) (10-shot, 96.66)}; 

\addplot [draw=purple, semithick, pattern=bricks,  pattern color = purple] coordinates {(Zero-Shot, 97.16) (2-shot, 98.49) (4-shot, 96.75) (6-shot, 99.32) (8-shot, 98.91) (10-shot, 99.99)}; 

\legend{C, C++, C\#, Python, Java}
\end{axis}
 
        \end{tikzpicture}
        \caption{Correct@1–5 averages for five programming languages across few-shot settings. }
         \label{perf}
    \end{figure*}

Table \ref{performance_comparison} presents the average Correct@1–5 scores across five programming languages—C, C++, C\#, Python, and Java—evaluated under varying shot configurations ranging from 0-shot to 10-shot prompting. Overall, the results reveal that all languages maintain consistently high correctness, typically surpassing 95\% accuracy regardless of the number of examples provided. Python exhibits the lowest performance under zero-shot prompting (78.66\%), indicating its stronger reliance on in-context examples compared to other languages. However, it rapidly improves once few-shot guidance is introduced, peaking at 99.32\% in the 2-shot setting. Java and C demonstrate exceptional stability, frequently exceeding 98–99\% correctness across multiple shot counts, while C\# achieves its highest accuracy with 4-shot prompting (99.32\%) before experiencing minor fluctuations. Interestingly, increasing shots does not always correlate with higher performance; certain languages achieve their peak correctness at mid-range shot levels (2–6 shots) rather than at the highest shot count. Figure \ref{perf} illustrates the correctness accuracy of various programming language in different shot settings. 
\begin{table}[h]
\caption{Average Correct@1–5 scores for five programming languages across different few-shot settings, illustrating the effect of shot count on code generation correctness.}
\label{performance_comparison}
\centering
\begin{tabular}{c|c|c|c|c|c}
\hline
Shots & C & C++ & C\# & Python & Java  \\
\hline \hline
0-shot & 97.66 & 97.41 & 95.63 & 78.66 & 97.16  \\ \hline  
2-shot & 96.07 & 96.32 & 97.49 & 99.32 & 98.49 \\  
\hline
4-shot & 97.88 & 96.38 & 99.32 & 98.66 & 96.75  \\   
\hline
6-shot & 99.32 & 99.32 & 98.49 & 95.99 & 99.32  \\   
\hline
8-shot & 97.91 & 98.16 & 93.27 & 93.88 & 98.91  \\   
\hline
10-shot & 97.82 & 97.07 & 93.08 & 96.66 & 99.99  \\  \hline  

\end{tabular}
\end{table}

 Table \ref{distance} presents a comprehensive evaluation of Distance and Similarity metrics for refactored code. Distance measures the structural deviation between the refactored code and the original, while Similarity quantifies the degree of lexical or semantic alignment. Across all shot settings, Python consistently achieves the lowest mean distances, ranging approximately from 276.93 (2-shot) to 293.67 (10-shot), indicating that refactored Python code preserves structural fidelity better than other languages. In contrast, Java and C\# frequently exhibit the highest distances, particularly at higher shots, reaching up to 509.93 for Java at 6-shot that reflects more substantial structural modifications during refactoring. C and C++ generally maintain moderate distances, suggesting that while the structural changes are noticeable, they are not as extreme as those observed for Java or C\#. In terms of similarity, C\# consistently achieves the highest mean values, peaking at 57.63\% under 10-shot prompting that indicates that despite larger structural changes, the semantic and lexical content is largely retained. Java also maintains high similarity values at higher shots ($\approx$53–54\%) that highlights its ability to balance structural changes with semantic preservation. Python exhibits moderate similarity values ($\approx$44–48\%), which, combined with its low distance, that indicates that the refactoring is consistent and minimally disruptive. C and C++ display variable similarity across shots, with mean values ranging from 36\% to 55\% that reflects less predictable outcomes depending on the shot configuration. Examining shot-wise trends, low-shot settings (0–2 shots) generally result in higher distances and lower similarity for most languages which suggests that minimal examples are insufficient to guide the model effectively. Moderate-shot settings (4–6 shots) tend to improve both metrics and achieve a better balance between structural fidelity and semantic preservation. At high-shot settings (8–10 shots), Python maintains low distance, while C\# and Java continue to achieve high similarity which highlights the optimal shot configuration depends on the language and desired outcome. Overall, Python is the most structurally consistent across all shots, C\# preserves semantic content most effectively, and Java provides a balanced trade-off between structural modification and semantic similarity at higher shots, emphasizing the importance of language-specific strategies and shot configurations in automated code refactoring.

\begin{table*}[!t]
\caption{Distance and Similarity metrics for refactored code across five programming languages under varying shot settings, showing mean, standard deviation, and min–max ranges to capture structural and semantic changes.}
\label{distance}
\centering
\begin{tabular}{c|c|cc|cc}
\hline
\multirow{2}{*}{\textbf{Shots}} & \multirow{2}{*}{\textbf{Programming Language}} 
& \multicolumn{2}{c|}{\textbf{Distance}} & \multicolumn{2}{c}{\textbf{Similarity}} \\ 
 &  & Mean(Std) & $Min \sim Max$ & Mean(Std) & $Min \sim Max$ \\
 
 \hline \hline

\multirow{5}{*}{0-shot} 
& C      & 327.19 (± 20.74) & 228.00 ~ 400.00 & 45.10\% (± 3.48\%) & 32.89\% ~ 61.74\%  \\ \cline{2-6}
& C++    & 338.70 (± 23.44) & 262.00 ~ 497.00 & 45.19\% (± 3.79\%) & 19.58\% ~ 57.61\% \\ \cline{2-6}
& C\#    & 465.27 (± 40.03) & 335.00 ~ 607.00 & 40.12\% (± 5.15\%) & 21.88\% ~ 56.89\%  \\ \cline{2-6}
& Python & 294.92 (± 20.85) & 252.00 ~ 411.00 & 44.39\% (± 3.89\%) & 22.45\% ~ 52.45\%  \\ \cline{2-6}
& Java   & 484.62 (± 41.08) & 336.00 ~ 603.00 & 33.98\% (± 5.60\%) & 17.85\% ~ 54.22\%  \\ \hline

\multirow{5}{*}{2-shot} 
& C      & 327.99 (± 21.14) & 218.00 ~ 413.00 & 44.96\% (± 3.55\%) & 30.70\% ~ 63.42\%  \\ \cline{2-6}
& C++    & 321.71 (± 30.15) & 256.00 ~ 814.00 & 48.28\% (± 3.58\%) & 18.52\% ~ 58.58\%  \\ \cline{2-6}
& C\#    & 340.23 (± 30.90) & 258.00 ~ 525.00 & 56.21\% (± 3.98\%) & 32.43\% ~ 66.80\%  \\ \cline{2-6}
& Python & 276.93 (± 22.39) & 243.00 ~ 371.00 & 47.75\% (± 4.22\%) & 30.00\% ~ 54.15\%  \\ \cline{2-6}
& Java   & 366.52 (± 35.86) & 297.00 ~ 541.00 & 50.13\% (± 4.90\%) & 26.29\% ~ 59.54\%  \\ \hline

\multirow{5}{*}{4-shot} 
& C      & 303.27 (± 20.76) & 229.00 ~ 387.00 & 49.12\% (± 3.48\%) & 35.07\% ~ 61.58\%  \\ \cline{2-6}
& C++    & 298.04 (± 17.27) & 259.00 ~ 460.00 & 51.77\% (± 2.79\%) & 25.57\% ~ 58.09\%  \\ \cline{2-6}
& C\#    & 336.52 (± 29.40) & 234.00 ~ 639.00 & 56.72\% (± 3.80\%) & 17.76\% ~ 69.88\%  \\ \cline{2-6}
& Python & 293.85 (± 19.29) & 240.00 ~ 384.00 & 44.57\% (± 3.61\%) & 30.05\% ~ 54.72\%  \\ \cline{2-6}
& Java   & 381.26 (± 56.08) & 283.00 ~ 585.00 & 48.06\% (± 7.64\%) & 20.30\% ~ 61.44\%  \\ \hline

\multirow{5}{*}{6-shot} 
& C      & 278.35 (± 32.28) & 223.00 ~ 439.00 & 53.30\% (± 5.41\%) & 26.34\% ~ 62.58\%  \\ \cline{2-6}
& C++    & 420.29 (± 18.11) & 231.00 ~ 485.00 & 31.99\% (± 2.93\%) & 21.52\% ~ 62.62\%  \\ \cline{2-6}
& C\#    & 337.33 (± 40.19) & 235.00 ~ 564.00 & 56.59\% (± 5.17\%) & 27.41\% ~ 69.76\%  \\ \cline{2-6}
& Python & 292.47 (± 30.15) & 229.00 ~ 435.00 & 44.81\% (± 5.69\%) & 17.92\% ~ 56.79\%  \\ \cline{2-6}
& Java   & 509.93 (± 22.27) & 350.00 ~ 552.00 & 30.53\% (± 3.03\%) & 24.80\% ~ 52.32\%  \\ \hline

\multirow{5}{*}{8-shot} 
& C      & 266.33 (± 29.19) & 206.00 ~ 450.00 &  55.34\% (± 4.93\%) & 24.50\% ~ 68.50\%  \\ \cline{2-6}
& C++    & 395.13 (± 50.76) & 229.00 ~ 466.00 & 36.06\% (± 8.21\%) & 24.60\% ~ 62.94\%  \\ \cline{2-6}
& C\#    & 384.07 (± 46.00) & 269.00 ~ 643.00 & 50.57\% (± 5.92\%) & 17.25\% ~ 65.38\%  \\ \cline{2-6}
& Python & 301.74 (± 26.39) & 192.00 ~ 401.00 & 43.07\% (± 4.98\%) & 24.34\% ~ 63.77\%  \\ \cline{2-6}
& Java   & 339.45 (± 16.89) & 303.00 ~ 433.00 & 53.75\% (± 2.30\%) & 41.01\% ~ 58.72\%  \\ \hline

\multirow{5}{*}{10-shot} 
& C      & 269.11 (± 23.97) & 192.00 ~ 476.00 &  54.85\% (± 4.02\%) & 20.13\% ~ 67.79\%  \\ \cline{2-6}
& C++    & 306.83 (± 32.44) & 234.00 ~ 439.00 & 50.35\% (± 5.25\%) & 28.96\% ~ 62.14\%  \\ \cline{2-6}
& C\#    & 329.25 (± 44.95) & 204.00 ~ 665.00 & 57.63\% (± 5.78\%) & 14.41\% ~ 73.75\%  \\ \cline{2-6}
& Python & 293.67 (± 21.15) & 240.00 ~ 421.00 & 44.64\% (± 3.92\%) & 20.57\% ~ 54.72\%  \\ \cline{2-6}
& Java   & 339.33 (± 26.12) & 255.00 ~ 524.00 & 53.77\% (± 3.56\%) & 28.61\% ~ 65.26\%  \\ \hline

\end{tabular}
\end{table*}

\section{Conclusion}
This study presents a comprehensive investigation into the effectiveness of LLMs for multilingual code refactoring across C, C++, C\#, Python, and Java. By designing a prompting module with ten distinct prompts and employing zero-shot, few-shot, and multi-shot strategies, we fine-tuned the model to generate high-quality refactored code while preserving semantic correctness. Quantitative evaluation across multiple metrics including compilability, cyclomatic complexity, lines of code, character and token counts that demonstrates the model’s ability to produce syntactically correct and maintainable code with notable improvements in structural simplicity. The Results show that Java stands out as the top-performing language, achieving not only the highest average compilability rate of 94.78\% but also the best correctness score of 99.99\% in the 10-shot setting, This demonstrates that the model’s strong reliability when provided with sufficient contextual examples. Python, on the other hand, consistently yields the lowest cyclomatic complexity and suggests that the refactored outputs tend toward simpler and more maintainable structures. Across all shot configurations, the Correct@k performance remains consistently high across languages, indicating that even minimal prompting is sufficient for generating functionally accurate code. The distance and similarity analyses further confirm that the refactored programs preserve core semantic behavior while introducing meaningful structural improvements. Additionally, the comparative results between original and LLM-refactored implementations clearly show that the model is capable of reducing code complexity and lines of code without sacrificing functional integrity.

%\section*{Acknowledgment}
%This research received financial support from the Japan Society for the Promotion of Science (JSPS) KAKENHI, with grant number 23H03508.

\bibliographystyle{IEEEtran}

\vfill

\end{document}